%% file: attention2d.tex
\documentclass[sigconf,natbib=true,anonymous=false]{acmart}

\AtBeginDocument{%
  \providecommand\BibTeX{{%
    \normalfont B\kern-0.5em{\scshape i\kern-0.25em b}\kern-0.8em\TeX}}}

\setcopyright{acmcopyright}
\copyrightyear{2022}
\acmYear{2022}

\acmConference[WSDM '22]{WSDM '22: The 15th ACM  International Conference on Web Search and Data Mining}{February 21--25, 2022}{Phoenix, AZ, USA}
\acmBooktitle{WSDM '22: The 15th ACM  International Conference on Web Search and Data Mining,
  February 21--25, 2022, Phoenix, AZ, USA}
\acmPrice{15.00}

\graphicspath{ {Figures/} }
\usepackage[ruled,vlined]{algorithm2e}  
\usepackage{varwidth}
\usepackage{amsfonts}
\usepackage{xspace}
\usepackage{enumitem}
\usepackage{array}
\usepackage{multirow}
\usepackage{caption}
\usepackage{subcaption}
\usepackage{tablefootnote}
\usepackage[export]{adjustbox}
\usepackage{xcolor}
\usepackage{amsthm}
\usepackage{stackengine}
\usepackage{amsmath}
\usepackage{makecell}
\usepackage{comment}
\usepackage{url}
\usepackage{balance}

\begin{document}


\title{Sequential Modeling with Multiple Attributes for Watchlist Recommendation in E-Commerce}


\renewcommand{\thefootnote}{\fnsymbol{footnote}}

\author{Uriel Singer$^1$, Haggai Roitman$^1$, Yotam Eshel$^1$, Alexander Nus$^1$, Ido Guy$^{2,}$\footnotemark\footnotetext[1]{Work was done while still at eBay}, Or Levi$^1$, Idan Hasson$^1$, Eliyahu Kiperwasser$^1$}
\affiliation{%
  \institution{$^1$eBay Research, Israel}
  \country{}
}
\email{{usinger,hroitman,yeshel,alnus,olevi,ihasson,ekiperwasser}@ebay.com}

\affiliation{%
  \institution{$^2$Ben-Gurion University of the Negev, Israel}
  \country{}
}
\email{idoguy@acm.com}




\renewcommand{\shortauthors}{Singer et al.}

\input{keywords}


\input{macros}
\begin{abstract}
\input{0-abstract}
\end{abstract}

\maketitle
\renewcommand{\thefootnote}{\arabic{footnote}}
\input{1-intro}

\input{2-rw}

\input{3-framework}
\input{4-setup}
\input{4-results}

\input{5-conclusions}


\balance
\bibliographystyle{ACM-Reference-Format}
\bibliography{attention2d}


\end{document}

%% file: keywords.tex
\ccsdesc[500]{Applied computing~Electronic commerce}

\keywords{Watchlist, Sequential-Model, Transformers, E-Commerce}

%% file: macros.tex
\newcommand{\specialcell}[2][c]{%
  \begin{tabular}[#1]{@{}c@{}}#2\end{tabular}}

\def\showcomments{}
\ifdef{\showcomments}
{
\newcommand{\us}[1]{\textcolor{green}{$\ll$\textsf{#1 --US}$\gg$}}
\newcommand{\ye}[1]{\textcolor{purple}{$\ll$\textsf{#1 --YE}$\gg$}}
\newcommand{\hr}[1]{\textcolor{blue}{$\ll$\textsf{#1 --HR}$\gg$}}
\newcommand{\ol}[1]{\textcolor{red}{$\ll$\textsf{#1 --OL}$\gg$}}
\newcommand{\an}[1]{\textcolor{brown}{$\ll$\textsf{#1 --AN}$\gg$}}
\newcommand{\ih}[1]{\textcolor{yellow}{$\ll$\textsf{#1 --IH}$\gg$}}
\newcommand{\ig}[1]{\textcolor{pink}{$\ll$\textsf{#1 --IG}$\gg$}}
\newcommand{\ek}[1]{\textcolor{orange}{$\ll$\textsf{#1 --EK}$\gg$}}
}
{
\newcommand{\us}[1]{}
\newcommand{\ye}[1]{}
\newcommand{\hr}[1]{}
\newcommand{\ol}[1]{}
\newcommand{\an}[1]{}
\newcommand{\ih}[1]{}
\newcommand{\ig}[1]{}
\newcommand{\ek}[1]{}
}

\newcommand\footnoteref[1]{\protected@xdef\@thefnmark{\ref{#1}}\@footnotemark}
\newcommand\oast{\stackMath\mathbin{\stackinset{c}{0ex}{c}{0ex}{\ast}{\bigcirc}}}
\SetKwInOut{KwInput}{Input} 
\SetKwInOut{KwOutput}{Output}

\newcommand{\wl}{{watchlist}\xspace}
\newcommand{\wls}{{watchlists}\xspace}
\newcommand{\Atd}{{Attention2D}\xspace}
\newcommand{\Ttd}{{Trans2D}\xspace}
\newcommand{\atd}{{attention2D}\xspace}
\newcommand{\Lod}{{Linear1D}\xspace}
\newcommand{\Ltd}{{Linear2D}\xspace}
\newcommand{\MHtd}{{MultiHead2D}\xspace}
\newcommand{\td}{{tensordot}\xspace}

\newcommand{\Ltdb}{{Linear2D^+}\xspace}
\newcommand{\AtdH}{{ScaledDotProductAttention2D}\xspace}

\newcommand{\argmin}{\mathop{\mathrm{argmin}}} 

\newcolumntype{L}[1]{>{\raggedright\let\newline\\\arraybackslash\hspace{0pt}}m{#1}}
\newcolumntype{C}[1]{>{\centering\let\newline\\\arraybackslash\hspace{0pt}}m{#1}}
\newcolumntype{R}[1]{>{\raggedleft\let\newline\\\arraybackslash\hspace{0pt}}m{#1}}
\newcommand\independent{\protect\mathpalette{\protect\independenT}{\perp}}
\def\independenT#1#2{\mathrel{\rlap{$#1#2$}\mkern2mu{#1#2}}}
\def\multitable#1{\multicolumn{1}{p{0.5cm}}{\centering #1}}

\newcommand{\cdmf}{{\textbf{CDMF}}\xspace}
\newcommand{\seqfm}{{\textbf{SeqFM}}\xspace}
\newcommand{\fdsa}{{\textbf{FDSA}}\xspace}
\newcommand{\gru}{{\textbf{GRU}}\xspace}
\newcommand{\trans}{{\textbf{Trans}}\xspace}
\newcommand{\lc}{{\textbf{LatentCross}}\xspace}
\newcommand{\ca}{{\textbf{ContextAware}}\xspace}
\newcommand{\bst}{{\textbf{BST}}\xspace}
\newcommand{\xgb}{{\textbf{Production}}\xspace}
\newcommand{\RSP}{{\textbf{RSP}}\xspace}
\newcommand{\price}{{\textbf{Price}}\xspace}
\newcommand{\seept}{{\textbf{SSE-PT}}\xspace}

\newcommand{\grurec}{{\textbf{GRU4Rec}}\xspace}
\newcommand{\bertrec}{{\textbf{BERT4Rec}}\xspace}
\newcommand{\sasrec}{{\textbf{SASRec}}\xspace}

\def\PP{{\mathbb P}}
\def\EE{{\mathbb E}}
\def\RR{{\mathbb R}}
\def\cov{\text{cov}}

\def\U{\mathbf{U}}
\def\WL{\mathbf{WL}}
\def\VI{\mathbf{VI}}
\def\V{\mathbf{V}}
\def\S{\mathbf{S}}
\def\W{\mathbf{W}}
\def\K{\mathbf{K}}
\def\Q{\mathbf{Q}}
\def\I{\mathbf{I}}
\def\A{\mathbf{A}}
\def\P{\mathbf{P}}
\def\X{\mathbf{X}}
\def\Y{\mathbf{Y}}
\def\E{\mathbf{E}}
\def\H{\mathbf{H}}
\def\x{\mathbf{x}}
\def\q{\mathbf{q}}
\def\k{\mathbf{k}}
\def\v{\mathbf{v}}
\def\o{\mathbf{o}}
\def\e{\mathbf{e}}

%% file: 0-abstract.tex
In e-commerce, the \wl enables users to track items over time and has emerged as a primary feature, 
playing an important role in users' shopping journey. Watchlist items typically have multiple attributes whose values may change over time (e.g., price, quantity). Since many users accumulate dozens of items on their \wl, and since shopping intents change over time, recommending the top \wl items in a given context can be valuable. In this work, we study the \wl functionality in e-commerce and introduce a novel \wl recommendation task. Our goal is to prioritize which \wl items the user should pay attention to next by predicting the next items the user will click.  We cast this task as a specialized sequential recommendation task and discuss its characteristics. 
Our proposed recommendation model, \Ttd, is built on top of the Transformer architecture, where we further suggest a novel extended attention mechanism (\Atd) that allows to learn complex item-item, attribute-attribute and item-attribute patterns from sequential-data with multiple item attributes. 
Using a large-scale \wl dataset from eBay, 
we evaluate our proposed model, where we demonstrate its superiority compared to multiple state-of-the-art baselines, many of which are adapted for this task.
 

%% file: 1-intro.tex
\section{Introduction}
The \wl, a collection of items to track or follow over time, has become a prominent feature in a variety of online applications spanning multiple domains, including news reading, television watching, and stock trading. One domain in which watchlists have become especially popular is e-commerce, where they allow users to create personalized collections of items they consider to purchase and save them in their user account for future reference. Saving items in the watchlist allows users to track a variety of dynamic characteristics over relatively long period of times. These include the  price and related deals, 
shipping costs, delivery options, etc. 


\subsection{The Watchlist Recommendation Task}

A user's \wl can be quite dynamic. At any moment, the user may choose to revise her \wl by adding new items (e.g., after viewing an item's page) or removing existing items (e.g., the user has lost her interest in an item). Over time, item attributes may change (e.g., an item's price has dropped) 
or items may become invalid for recommendation (e.g., an item has been sold out). 
User's latest shopping journey may change as well, as she may show interest in other items, categories or domains. 

All in all, in between two consecutive interactions with the \wl, the user may change her priority of which next item to pay attention to. 
Considering the fact that the \wl of any user may include dozens of items, it may be quite overwhelming for users to keep track of all changes and opportunities related to their \wl items. Given a limited number of items that can be displayed to the user (e.g., 2-3 items on eBay's mobile homepage watchlist module) and the dynamic nature of watchlist items, our goal is, therefore, to help users prioritize which watchlist items they should pay attention to next. 

We cast the  \wl recommendation (hereinafter abbreviated as WLR) task as a specialized \emph{sequential recommendation} (SR) task. Given the historical interactions of a user with items, the goal of the SR task is to predict the next item the user will interact with~\cite{wang2019sequential}.  
While both tasks are related to each other, we next identify two important characteristics of the WLR task. 

The first 
characteristic of the WLR task is that, at every possible moment in time, only a subset of items, \textbf{explicitly chosen} to be tracked by the user prior to recommendation time, should be considered. Moreover, the set of items in a given user's \wl may change from one recommendation time to another. Yet, most previous SR studies assume that, the user's next interaction may be with \textbf{any possible} item in the catalog. Therefore, the main focus in such works is to predict the next item's \textbf{identity}. 
The WLR task, on the other hand, aims to estimate the user's attention to \wl items by predicting the \textbf{click likelihood} of each item in the user's \wl. Noting that \wl datasets may be highly sparse, as in our case, representing items solely by their identity usually leads to over-fit. A better alternative is to represent items by their attributes. 
Yet, to the best of our knowledge, only few previous SR works can consider attribute-rich inputs both during training and prediction. 

The second characteristic of the task lies in the important observation that, possible shifts in the user's preferences towards her watched items may be implied by her recently viewed items (RVIs). For example, a user that tracks a given item in her \wl, may explore alternative items (possibly not in her \wl) from the same seller or category, prior to her decision to interact with the tracked item (e.g., the watched item price is more attractive). 
Trying to handle the unique challenges imposed by the WLR task, we design an extension to the Transformer model~\cite{vaswani2017attention}. 
Our proposed Transformer model, \textbf{\Ttd}, employs a novel attention mechanism (termed \emph{\Atd}) that allows to learn complex preferential patterns from historical sequences of user-item interactions accompanied with a rich and dynamic set of attributes. The proposed attention mechanism is designed to be highly effective, and 
requires only a small addition to the model's parameters. 

Using a large-scale user \wl dataset from eBay, we evaluate the proposed model. We show that, by employing the novel \Atd mechanism within the Transformer model, our model can achieve superior recommendation performance, compared to a multitude of state-of-the-art SR models, most of which we adapt to this unique task.

%% file: 2-rw.tex
\section{Related Work}
\label{sec:rw}
The \wl recommendation (WLR) task can be basically viewed as a specialization of the sequential recommendation (SR) task. 
The majority of previous SR works assume that the input sequence includes only item-IDs~\cite{wang2019sequential} 
and focus on predicting the identity of the next item that the user may interact with~\cite{wang2019sequential}. 
Notable works span from Markov-Chain (MC)~\cite{10.1145/1772690.1772773} and translation-based~\cite{10.1145/3109859.3109882} models that aim to capture high-order relationships between users and items, to works that employ various deep-learning models that capture user's evolving preferences such as RNNs (e.g., GRU~\cite{hidasi2016sessionbased, 10.1145/2988450.2988452}), CNNs (e.g.,~\cite{10.1145/3159652.3159656,10.1145/3109859.3109900}), and more recent works that utilize Transformers (e.g., SASRec~\cite{kang2018self},  BERT4Rec~\cite{sun2019bert4rec} and SSE-PT~\cite{10.1145/3383313.3412258}). 

Similar to~\cite{kang2018self,sun2019bert4rec,10.1145/3383313.3412258}, we also utilize the Transformer architecture. Yet, compared to existing works that train models to predict the next item-ID, our model is trained to rank \wl items according to their click likelihood.
The main disadvantage of models that focus on item-ID only inputs is their inability to handle datasets having items with high sparsity and item cold-start issues. Utilizing rich item attributes, therefore, becomes extremely essential for overcoming such limitations, and specifically in our task.
By sharing attribute-representations among items, such an approach can better handle sparse data and item cold-start. 

Various approaches for obtaining rich attribute-based item representations during \textbf{training-time} have been studied so far. This includes a variety of attribute representations aggregation methods~\cite{mizrachi2019combining,10.1145/3159652.3159727,10.1145/3121050.3121061,sar2018collaborative} (e.g., average, sum, element-wise multiplication, etc.),  concatenation~\cite{mizrachi2019combining,zhou2018atrank,10.1145/3326937.3341261}, parallel processing~\cite{10.1145/2959100.2959167,10.1145/3397271.3401111}, usage of attention-based Transformer layer~\cite{zhang2019feature} and graph-based representations~\cite{wang2020knowledge}. Furthermore, attributes have been utilized for introducing additional mask-based pre-training tasks~\cite{zhou2020s3} (e.g., predict an item's attribute).

Yet, during prediction, leveraging the attributes is not trivial, as the entire catalog should be considered. Indeed,  all aforementioned works still utilize only item-IDs for prediction.
For example, the S3~\cite{zhou2020s3} model, one of the leading self-supervised SR models, supports only static attributes, which are only considered during its pre-training phase. 
Hence, extending existing works for handling the WLR task, which requires to consider also item attributes at prediction time, is extremely difficult, and in some cases, is just impossible. As we shall demonstrate in our evaluation (Section~\ref{sec:experiments}), applying those previous works~\cite{mizrachi2019combining,sar2018collaborative,10.1145/3326937.3341261,zhang2019feature} that can be extended to handle the WLR task (some with a lot of effort and modification), results in inferior recommendation performance compared to our proposed model. 

Few related works allow to include item attributes as an additional context that can be utilized for representing items both during training and  prediction time. 
Such works either 
extend basic deep-sequential models (RNNs~\cite{10.1145/3159652.3159727,smirnova2017contextual} and Transformers~\cite{eshel2021presize}) with context that encodes attribute-rich features,
or combine item-context with sequential data using Factorization Machines~\cite{pasricha2018translation,chen2020sequence}. 

Our proposed model is also capable of processing sequential inputs having multiple item attributes. To this end, our model employs a modified Transformer architecture with a novel attention mechanism (\Atd) that allows to process complete sequences of 2D-arrays of data without the need to pre-process items (e.g., aggregate over attributes) in the embedding layer (as still need to be done in all previous works). This allows our model, with a slight expense of parameters, to capture highly complex and diverse preference signals, 
which are preserved until the prediction phase.

%% file: 3-framework.tex
\section{Recommendation Framework}
\label{sec:framework}

\subsection{Problem Formulation}
\label{sec:formulation}
The WLR task can be modeled as a special case of the sequential recommendation (SR) task~\cite{wang2019sequential}, with the distinction that, at every possible moment, there is only a limited, yet dynamic, valid set of watched items that need to be considered.  Using the user's historical interactions with \wl items and her recently viewed item pages (RVIs), our goal is, therefore, to predict which items in the current user's \wl will be clicked. A user's click on a \wl item usually implies that the user has regained her interest in that item. Moreover, we wish to leverage the fact that watched items are dynamic, having attributes whose values may change over time (e.g., price), and therefore, may imply on a change in user's preferences. 

Formally, 
let $\I$ be the set of all recommendable items. 
For a given user $u$, 
let $\WL_t\subseteq{\I}$ denote the set of items in the user's \wl \textbf{prior} to the time of $t$-th \textbf{click} on any of the \wl items, termed also hereinafter a ``\emph{\wl-snapshot}''. We further denote $m_t=|\WL_t|$ the number of items in a given snapshot. A user may click on any item in $\WL_t$, which leads her to the item's page; also termed a ``view item'' (VI) event.   
VI events may also occur \textbf{independently} of items in the user's \wl; e.g., the user views an  item in $\I$ following some search result or directed to the item's page through some other organic recommendation module, etc. Let $\VI_t\subseteq{\I}$, with $n_t=|\VI_t|$, denote a possible sequence of item pages viewed by user $u$ between the $t-1$-th and $t$-th clicks on her \wl items, further termed a ``\emph{view-snapshot}''.  

We next denote $\S(t)$ -- the user's history prior to her $t$-th interaction with the \wl. 
$\S(t)$ is given by the sequence of the user's previous clicks on \wl items and items whose page the user has viewed in between each pair of consecutive clicks, i.e.: 
\vspace{-0.05in}
\[
\S(t)=\underbrace{w_{1}}_{click},
\underbrace{v_{1,1},\ldots,v_{1,n_1}}_{\VI_{1}},
\ldots,\underbrace{w_{t-1}}_{click},
\underbrace{v_{t-1,1},\ldots,v_{t-1,n_{t-1}}}_{\VI_{t-1}}.
\]

Here, $w$ denotes a single user's click on some \wl item, while $v$ denotes an item page viewed by the user. We note that, every \wl item clicked by the user leads to the item's page, hence: $w_{l}=v_{l,1}$ ($\forall{l}:1\leq{l}\leq{t-1}$).  


Using the above definitions, for any given user history $\S(t)$ and new \wl snapshot $\WL_{t}$ to consider, our goal is to predict which items in $\WL_{t}$ the user is mostly likely to click next. Therefore, at service time, we wish to recommend to the user the top-$k$ items in $\WL_{t}$ with the highest click likelihood.

\subsection{Recommendation Model}\label{sec:model}

\begin{figure}
\centering
\includegraphics[width=0.5\textwidth]{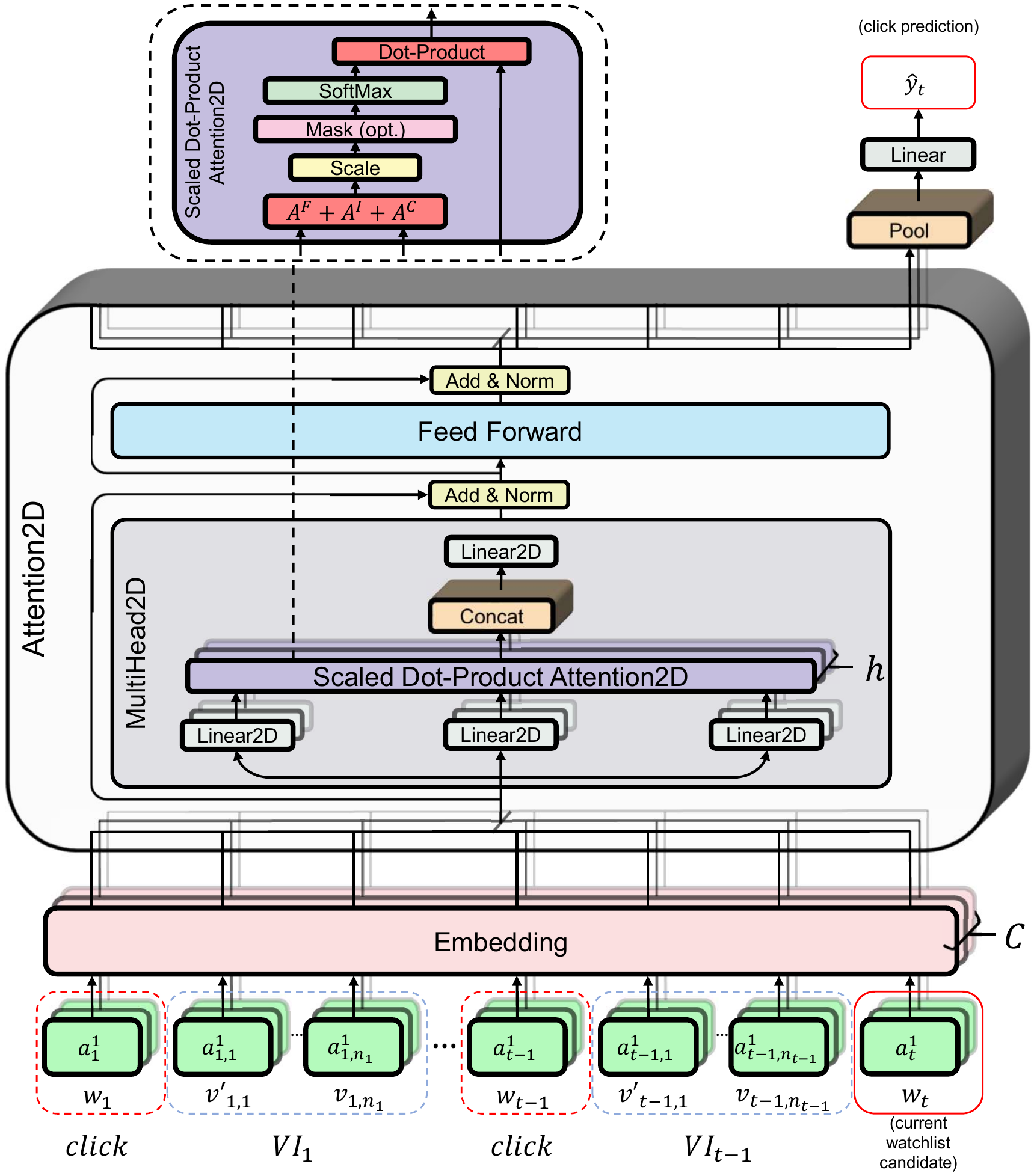}
\caption{Watchlist recommendation model (\Ttd) using Transformer with \Atd layer. The model has three main parts: Embedding layer (bottom-side), Attention2D layer (middle-side) and Prediction layer (upper-right side) The ScaledDotProductAttention2D component is further illustrated in the upper-left side.
}
\label{fig:model}
\end{figure}

Our WLR model, \textbf{\Ttd}, extends the Transformer~\cite{vaswani2017attention} architecture for handling sequences with items that include several categorical attributes as in the case of \wl items. Using such an architecture allows us to capture high-order dependencies among items and their attributes with a minimum loss of semantics and a reasonable model complexity. The model's network architecture is depicted in Figure~\ref{fig:model}. The network includes three main layers, namely, \emph{Embedding} layer, \emph{\Atd} layer and \emph{Prediction} layer. We next describe the details of each layer.

\subsection{Embedding Layer}
\label{sec:embedding}
 A common modeling choice in most related literature~\cite{wang2019sequential}, is to represent user's sequence $\S(t)$ as the sequence of item-IDs. 
Let $N=\sum_{l=1}^{t-1}(1+n_{l})$ denote the total items in $\S(t)$. These item-IDs are typically embedded into a corresponding sequence of $d$-dimensional vectors $\E = (\e_1,\ldots,\e_N); \forall{i}:\e_i \in \RR^{d} $ using an embedding dictionary and serve as input representation to neural-network models. 
However, such a representation has two main disadvantages. Firstly, it is prone to cold-start problems as items that never appear during training must be discarded or embedded as an `unknown' item. Secondly, items that only appear once or a few times during training are at high risk of over-fitting by the training process memorizing their spurious labels.
Such sparsity issues are particularly present in our data (as will be discussed in Section~\ref{sec:dataset}), prompting us to represent an item as a collection of attributes, rather than an item-ID. Such an approach is less prone to sparsity and over-fitting, as the model learns to generalize based on item-attributes.

Following our choice for item representation, we now assume that each item in 
$\S(t)$ is represented by $C$ attributes: $[a_1,\ldots,a_C]$. Therefore, the input to our model is a 2D-array of item attributes $\S(t) = [a]_{i,j}$; where  $i \in [1,\ldots,N], j \in [1,\ldots,C]$. 
Hereinafter, we refer to the first axis of $\S(t)$ as the \textit{sequence} (item) dimension and its second axis as the \textit{channel} (attribute) dimension. 

The first layer of our model is an embedding layer with keys for all possible attribute values in our data. Therefore, by applying this layer on the user's sequence $\S(t)$, we obtain a 2D-array of embedding vectors $\E = [e]_{i, j} = Embd([a]_{i, j})$. Note that, since each item attribute embedding by itself is a $d$-dimensional vector, $\E\in\RR^{N\times{C}\times{d}}$ is 
a 3D-array. $\E$ is next fed into a specially \emph{modified Transformer} layer which is our model's main processing layer.

\subsection{Modified Transformer}
\label{sec:modified_transformer}
Most existing SR models (e.g.,~\cite{hidasi2016sessionbased,10.1145/3159652.3159656,kang2018self,sun2019bert4rec}) take a sequence of vectors as input. 
However, in our case, we do not have a 1D-sequence of vectors but rather a 2D-array representing a sequence of ordered attribute collections. As Transformers were shown to demonstrate state-of-the-art performance in handling sequential data in general~\cite{vaswani2017attention} and sequential recommendation data specifically \cite{kang2018self, sun2019bert4rec, zhou2020s3,10.1145/3383313.3412258,eshel2021presize}, we choose to employ Transformers and extend the \emph{Attention Mechanism} of the Transformer to handle 2D-input sequences, rather than a 1D-input sequence.

Since the vanilla attention model employed by existing Transformer architectures cannot be directly used for 2D-data, different pooling techniques were devised to reduce the channel dimension and produce a 1D-sequence as an input for the Transformer. For example, in BERT \cite{kenton2019bert},  token, position and sentence embeddings are summed before being fed into the Transformer Encoder. Other alternatives to summation were explored such as averaging, concatenation, or a secondary Transformer layer (e.g.,~\cite{10.1145/3326937.3341261,zhang2019feature}). Yet, most of these approaches suffer from a reduction in data representation happening prior to applying the attention workhorse. This makes it harder for the attention mechanism to learn rich attention patterns. 
An exception is the concatenation approach that sidesteps this problem but produces very long vectors, which in turn, requires Transformer layers to have a large input dimension making them needlessly large.
As an alternative, we next propose \textbf{\Atd} -- a dedicated novel attention block that is natively able to handle 2D-inputs in an efficient way. Such an extended attention mechanism will allow us to learn better attention patterns without significantly increasing the architecture size.

\subsection{\Atd}
\label{sec:attention2d}

    
We now describe in detail our modified attention block, referred to as \Atd. For comparison to the vanilla Transformer model, we encourage the reader to refer to~\cite{vaswani2017attention}.
An attention block maps a query ($\Q$) and a set of key ($\K$) - value ($\V$) pairs to an output, where the query, keys, values, and output are all vectors \cite{vaswani2017attention}. 
The \Atd block in turn receives a 2D-array of vectors, representing a sequence of ordered item attribute collections and computes the attention between all item-attributes while enabling different semantics per channel to better capture both channel interactions and differences.
This way, one attribute of a specific item can influence another attribute of a different item regardless of other attributes, making it possible to learn high-order dependencies between different items and attributes. For instance, in our usecase, the price preferences of a user, captured by the price channel can be computed using attention both to previously clicked item prices and to past viewed items of sellers correlated with pricey or cheap items. 
The full \Atd block is illustrated in  Figure~\ref{fig:model}. We now provide a detailed description of its implementation. 

\subsubsection{\Ltd}
\label{sec:fc2d}
The input to our model is a 2D-array of input vectors, generally denoted hereinafter $\X = [\x]_{i,j}$ (and specifically in our model's input $\X=\E$). Our model requires operations acting on such 2D-arrays throughout. To this end, we first define two extensions of standard linear neural-network layers:
\begin{equation}
\begin{split}
\Ltd_{W}(\X) = \W_j [\x]_{i,j} \\
\Ltdb_{W,b}(\X) = \W_j [\x]_{i,j} + b_j,
\end{split}
\end{equation}

where $\W = [W_1,\ldots,W_C]; b = [b_1,\ldots,b_C]$ and $\W_j \in \RR^{d \times d}; b_j \in \RR^{d}$ are trainable parameters per channel $1\leq j\leq C$. These operations define a linear layer (with or without bias) with different trainable parameters per channel and shall allow the model to facilitate interactions between different channels while preserving the unique semantics of each channel.  

We start by mapping our input ($\E$) into three 2D-arrays of query $\Q = [\q]_{i,j}$, key $\K = [\k]_{i,j}$ and value $\V = [\v]_{i,j}$ vectors corresponding to each input. We do so by applying three \Ltd layers:
\vspace{-0.12in}
$$\Q = \Ltd_{Q}(\E) \ ; \ \K = \Ltd_{K}(\E) \ ; \ \V = \Ltd_{V}(\E)$$

\subsubsection{\AtdH}
\label{sec:tensordot}
We next describe the core part of our proposed attention mechanism, which is used to compute a transformed representation of our 2D-input array. This part is further illustrated on the upper-left side of Figure~\ref{fig:model}. For brevity of explanation, we consider a single query vector $[\q]_{i,j}\in{\Q}$ and key vector $[\k]_{i',j'}\in{\K}$, while in practice, this operation is performed for every $i,i' \in [1,\ldots,N]$ and $j,j' \in [1,\ldots,C]$.

First, we compute the following three attention scores (where $T$ here denotes the matrix transpose operation): 
\vspace{-0.08in}
$$\A_{i, j, i', j'}^{F} = [\q]_{i, j} \cdot [\k]^T_{i', j'}\ ;\ \
\A_{i,i'}^{I} = \sum_{z=1}^C [\q]_{i, z} \cdot [\k]^T_{i', z}\ ;$$
\vspace{-0.05in}
$$\A_{j,j'}^{C} = \sum_{z=1}^N [\q]_{z, j} \cdot [\k]^T_{z, j'}$$ 

where:

$\bullet$ \textbf{$\A^F$} is a 4D-array of attention scores between all inputs as the dot-product between every query vector and every key vector. This attention corresponds to the attention scores in a vanilla attention layer between any two tokens, yet with the distinction that in our case we have 2D-inputs.

$\bullet$ \textbf{$\A^I$} 
 is a matrix of attention scores between whole items in our input, computed as the marginal attention over all channels. This attention captures the importance of items to each other regardless of a particular channel. 

$\bullet$ \textbf{$\A^C$}
is a matrix of attention scores between channels in our input, computed as the marginal attention over all items. This attention captures the importance of channels to each other regardless of a particular item.


The attention scores are further combined using their weighted sum:
\vspace{-0.15in}
\begin{equation}
\label{eq:final_attn}
\A_{i, j, i', j'} = \alpha_1 * \A^F_{i, j, i', j'} +  \alpha_2 * \A^I_{i,i'} + \alpha_3 * \A_{j,j'}^{C},
\end{equation}
where $\alpha_1,\alpha_2,\alpha_3$ are learned scalars, denoting the relative importance of each attention variant, respectively.

The result of this step is, therefore, a 4D-array $\A$ of attention scores from any position $(i,j)$ to every position $(i',j')$. Similarly to \cite{vaswani2017attention},
we apply a softmax function over $\A$,\footnote{For that, we flatten the last two dimensions and treat it as one long vector.} with a scaling factor of $\sqrt{d}$, so that our scores will sum up to 1.  We then compute the final transformed output $[\o]_{i,j}$ as a weighted average over all value vectors using the computed attention scores:
\vspace{-0.03in}
$$\P = softmax\left(\frac{\A}{\sqrt{d}}\right) \quad ; \quad [\o]_{i, j} = \sum_{i', j'} \P_{i, j, i', j'}[v]_{i', j'} $$

Finally, we can define our \Atd layer as an attention layer that receives a triplet of $(\Q, \K, \V)$ 2D-arrays of query, key and value vectors as its input and outputs a 2D-array transformed vectors:
\vspace{-0.03in}
\small
\begin{equation}
\begin{split}
\AtdH(\Q, \K, \V) = [\o]_{i, j} \\ i \in [1,\ldots,N], j \in [1,\ldots,C]
\end{split}
\end{equation}
\normalsize

The \Atd layer allows each item-attribute to attend to all possible item-attributes in the input. To prevent unwanted attendance (such as accessing future information in the sequence), we mask it out by forcing the relevant (future) attention values to $-\infty$ before the softmax operation.

\subsubsection{Multi-Head \Atd}
\label{sec:multihead}
Similarly to~\cite{vaswani2017attention}, instead of performing a single attention function, we can apply $h$ different functions (``Attention-Heads'') over different sub-spaces of the queries, keys, and values. This allows to diversify the attention patterns that can be learned, helping to boost performance. For instance, one attention head can learn to focus on price-seller patterns, while another on different price-condition patterns.
We facilitate this by applying $h$ different \Ltd layers, resulting in $h$ triplets of  $(\Q_r, \K_r, \V_r)$; on each, we apply the aforementioned \emph{\AtdH} layer. 
We then concatenate all $h$ results and project it back to a $d$-dimensional vector using another \Ltd layer:
\small
\begin{equation}
\begin{split}
\MHtd(\Q, \K, \V) = \Ltd_{W^O}(Concat(head_1,...,head_h)) \\
head_r = \AtdH(\Q_r, \K_r, \V_r) \\
\Q_r = \Ltd_{Q_r}(\Q) \ ; \ \K_r = \Ltd_{K_r}(\K) \ ; \ \V_r = \Ltd_{V_r}(\V) \\
\end{split}
\end{equation}
\normalsize

\subsubsection{Position-wise Feed-Forward Networks}
We finally allow the \Atd output to go through additional \Ltd layers in order to be able to cancel out unwanted applied alignments and learn additional complex representations:
\vspace{-0.03in}
$$FFN(\X) = \Ltdb_{W^2,b^2}(ReLU( \Ltdb_{W^1,b^1}(\X)))$$

\subsubsection{Full Layer implementation}
A full \Atd block is combined from a \MHtd layer and then a FFN layer, where a residual connection \cite{he2016deep} is applied around both, by adding the input to the output and applying a layer normalization \cite{ba2016layer}, as follows:
\vspace{-0.03in}
\begin{equation}\label{eq:trans}
\begin{split}
\Atd(\X)=LayerNorm(\X'+FFN(\X')), \\ \X'=LayerNorm(\X+\MHtd(\X, \X, \X))
\end{split}
\end{equation}

After applying the \Atd block, we end up with a transformed 2D-array of vectors $\E'=\Atd(\E)$, having the same shape as the original input. Therefore, \Atd layers can be further stacked one on the other to achieve deep attention networks.

\subsection{Prediction Layer}
The final stage in our model is label (click/no-click) prediction. Let $w_t\in{\WL_t}$ be an item in the user's \wl whose click likelihood we wish to predict. To this end, as a first step, we append $w_t$ to $\S(t)$. We then feed the extended sequence to the Transformer and obtain $\E'$, the transformed 2D-array representation based on our \Atd block (see again Eq~\ref{eq:trans}). 

We predict the click likelihood of item $w_t$ based on its transformed output $e'_{N+1}\in\RR^{d \times C}$. For that, we first obtain a single representation of it by applying $Pool(e'_{N+1})$ -- a pooling layer over the channel dimension in $e'_{N+1}$. While many pooling options may be considered, we use a simple average pooling. 
This is similar to the channel dimension reduction discussed in Section~\ref{sec:modified_transformer}, but instead of applying the pooling before the attention mechanism, we are able to apply the pooling after the \Atd block. This makes it possible for the \Atd block to capture better attention patterns between different items and attributes.

Finally, we obtain the item's predicted label (denoted $\hat{y}$) by applying a simple fully connected layer on $Pool(e'_{N+1})$ to a single neuron representing the item's click likelihood, as follows:
\vspace{-0.05in}
\begin{equation}
\label{eq:pred}
\hat{y}=\sigma(\W_p \cdot Pool(\e'_{N+1}) + b_p),
\end{equation}

where $\W_p \in \RR^{d \times 1}; b_p \in \RR$ are trainable parameters, and $\sigma(\cdot)$ is the Sigmoid activation function.

\subsection{Model Training and Inference}
To train the model, we first obtain a prediction for each item in the target \wl snapshot $\WL_{t}$ given the user's history $\S(t)$. Let $y\in\{0,1\}^{m_t}$ denote a binary vector having a single entry equal to $1$ for the actual item clicked in $\WL_{t}$ (and $0$ for the rest). We then train the model by applying the binary cross-entropy loss over this \wl snapshot: 
$Loss_{t} = \sum^{m_{t}}_{l=1}Loss_{l}$, where: $Loss_{l} = - \left[ {y_{l} log(\hat{y}_{l}) + (1-y_{l}) log(1-\hat{y}_{l})} \right]$; $\hat{y}_l$ is predicted according to Eq~\ref{eq:pred}.

At \textbf{inference}, we simply recommend the top-$k$ items in $\WL_{t}$ having the highest predicted click likelihood according to $\hat{y}$.

%% file: 4-setup.tex
\section{Evaluation}
\label{sec:experiments}

\subsection{Dataset}
\label{sec:dataset}
We collect a large-scale \wl dataset that was sampled from the eBay e-commerce platform during the first two weeks of February 2021. We further sampled only active users with at least 20 interactions with their \wl during this time period. Due to a sampling limit, for each \wl snapshot we are allowed to collect 
a maximum of 15 items, which are pre-ordered chronologically, according to the time they were added by the user to the \wl. 
In total, our dataset includes 40,344 users and 5,374,902 items. The data is highly sparse, having 11,667,759 and 1,373,794 item page views and \wl item clicks, respectively. An average \wl snapshot includes 10.48 items (stdev: 4.96). 
Items in our dataset hold multiple and diverse attributes. These include \emph{item-ID}, \emph{user-ID},  \emph{price} (with values binned into 100 equal sized bins using equalized histogram), \emph{seller-ID}, \emph{condition} (e.g., \texttt{new}, \texttt{used}), \emph{level1-category} (e.g., `\texttt{Building Toys}'), \emph{leaf-category} (e.g., `\texttt{Minifigures}'), \emph{sale-type} (e.g., \texttt{bid}) and \emph{site-ID} (e.g., \texttt{US}). Since item-IDs and seller-IDs in our dataset are quite sparse, we further hash these ids per sequence, based on the number of occurrences of each id type in the sequence. For each item in a given user's history, we keep several additional attributes, namely: \emph{position-ID} (similar to  BERT~\cite{kenton2019bert}), associated (\wl/view) \emph{snapshot-ID}, \emph{interaction type} (\texttt{\text{\wl}-click} or \texttt{item-page-view}), \emph{hour}, \emph{day}, and \emph{weekday} of interaction. Additionally, for each item in a given \wl snapshot, we keep its \emph{relative-snapshot-position} (\RSP) within the \wl (relative to user's inclusion time). We note that, we consider all position-based attributes relatively to the sequence end. 


 
 We split our data over time, having the train-set in the time range of 2--11 February 2021, and the test-set in the time range of 12--15 February 2021. We further use the most recent 1\% of the training time range as our validation-set. 

\subsection{Setup}
\subsubsection{Model implementation and training}
We implement \textbf{\Ttd}\footnote{GitHub repository with code and baselines: \url{https://github.com/urielsinger/Trans2D}} 
with pytorch~\cite{paszke2017automatic}. We use the Adam~\cite{kingma2015adam} optimizer, with a starting learning rate of $10^{-3}$, $\beta_1 = 0.9$, $\beta_2 = 0.999$, $\ell_2$ weight decay of $10^{-5}$, and dropout $p=0.3$.
 For a fair comparison, for all baselines (and our \textbf{\Ttd} model), we set the number of blocks $L=1$, number of heads $h=4$, embedding size $d=16$, and maximum sequence length $N=50$.
To avoid model overfitting, we add an exponential decay of the learning rate (i.e., at each epoch, starting from the second, we divide the current learning rate by $10$), and train for a total of 5 epochs.
We train all models on a single NVIDIA GeForce RTX 3090 GPU with a batch size of $32$. 

\subsubsection{Baselines}
\label{sec:baselines}
The WLR task requires to handle attribute-rich item inputs both during training and prediction. Yet, applying most existing baselines to directly handle the WLR task is not trivial. To recall, many previously suggested SR models support only inputs with item-IDs, and hence, are unsuitable for this task. Moreover, while many other SR models support attributes during training, their prediction is still based on item-IDs only. As the WLR task holds dynamic recall-sets, even those models that do support attributes during prediction, still require some form of adaptation to handle the WLR task. Overall, we suggest (and perform) three different model adaptations (denoted hereinafter \textbf{A1}, \textbf{A2} and \textbf{A3}, respectively) on such baselines, which we elaborate next. 


\textbf{A1}: Instead of using item-IDs for item embeddings, 
as most of our baselines do, 
we utilize their attributes. 
We obtain item input embeddings by averaging over each item's attribute embeddings. This adaptation always results in a better performance. Hence, we apply it to all baselines that do not naturally handle multi-attributes.


\textbf{A2}: Instead of predicting the next item, 
we train a model to predict clicks over items in the current \wl snapshot. This adaptation commonly helps to 
boost performance, as attributes of predicted items are utilized. Therefore, those baselines that do not offer a good alternative, are enhanced with this adaptation. We note that, predicting the next item over the current \wl snapshot only, 
always results in an inferior performance.  

\textbf{A3}: The \lc~\cite{10.1145/3159652.3159727} and \ca~\cite{smirnova2017contextual} baselines utilize context-features during prediction. 
The context representation is used as a mask to the GRU output. We adapt the context-features to be the average attribute embeddings of the \wl items.  

Having described the possible adaptations, we next list the set of baselines that we implement and compare to our proposed model. For each baseline, we further specify which adaptations we apply.  

$\bullet$  \RSP: Orders \wl items based solely on the attribute \emph{relative-snapshot-position}, i.e., relative to each item's user-inclusion time, having the newest item ranked first. 

$\bullet$  \price: Orders \wl items based solely on their \emph{price}. We report both descending and ascending price orderings. 

$\bullet$  $\gru_A$: Inspired by \grurec \cite{hidasi2016sessionbased}, we implement a GRU-based model, further applying adaptations \textbf{A1} and \textbf{A2}. 

$\bullet$  $\gru_C$: Similar to $\gru_A$, with the only change of concatenating the attribute embeddings instead of averaging them.

$\bullet$  $\trans_A$: Inspired by \bertrec \cite{sun2019bert4rec}, \sasrec \cite{kang2018self} and \seept \cite{10.1145/3383313.3412258}, we implement a Transformer-based model, which is further adapted using adaptations \textbf{A1} and \textbf{A2}.  

$\bullet$  $\bst$~\cite{10.1145/3326937.3341261}: Similar to $\trans_A$, with the only change of concatenating the attribute embeddings instead of averaging them.

$\bullet$  $\trans_T$: Inspired by \cite{eshel2021presize} and similar to $\trans_A$, with the only change of transforming the attribute embeddings 
by applying a vanilla-Transformer over the channel (attribute) dimension and only then averaging the embeddings.

$\bullet$  $\fdsa^+$: The original \fdsa~\cite{zhang2019feature} model offers two components: 1) Transformer over the item-IDs sequence, and 2) vanilla-attention pooling over the attribute embeddings followed by an additional Transformer over the attribute sequence. At the end, the two outputs are concatenated to predict the next item.
Inspired by \fdsa, we create a similar baseline but adapt its training to our task using adaptation \textbf{A2}. 
Here we note that, this baseline is the only one among all baselines that explicitly uses the \textbf{original} item-IDs.

$\bullet$  $\fdsa^-$: Similar to $\fdsa^+$, but without the item-ID Transformer; noting that using the original item-IDs (compared to their hashed version) may easily cause an over-fit over our dataset.

$\bullet$  \seqfm
~\cite{chen2020sequence}: 
Extends Factorization Machines~\cite{10.1145/2168752.2168771} (FM) with sequential data;  
Its inputs are a combination of static and sequence (temporal) features. 
It learns three different Transformers based on: 1) attention between the static features themselves 2) attention between the sequence features themselves, and 3) attention between the static and sequence features. It then pools the outputs of each Transformer into a single representation using average. Finally, it concatenates the three representations and predicts a given label. We adapt this baseline to our setting by first treating the predicted  \wl snapshot $\WL_t$ items' attributes as static features.  Additionally, applying adaptation \textbf{A1}, the dynamic features are obtained by averaging the attribute embeddings for all of the items in 
$\S(t)$. Using \textbf{A1}, 
this baseline can be directly used for click prediction.  

$\bullet$  \cdmf\footnote{\url{https://github.com/urielsinger/CDMF}}
~\cite{sar2018collaborative}: 
Implements an extended Matrix-Factorization (MF) method that handles complex data with multiple feedback types and repetitive user-item interactions. To this end, it receives as its input a sequence of all user-item (pair) interactions. The importance of an item to a given user is learned from an attention pooling over all the user-item interactions. The user representation is calculated as the weighted average over all her interacted items. The final prediction is calculated using the dot-product between the user and item representations.  Inspired by \cdmf, we implement the same architecture, where we apply adaptations \textbf{A1} and \textbf{A2}.  

$\bullet$  \lc~\cite{10.1145/3159652.3159727}: Incorporates context features over the output of the GRU layer which limits the prediction to specific context-features. Therefore, we apply adaptation \textbf{A3}, leveraging the context-features to be the next item's features. Using adaptation \textbf{A1}, item features are calculated as the average attribute embeddings.

$\bullet$  \ca~\cite{smirnova2017contextual}: Similar to \lc, with the exception that the context representation is treated both as a mask to the GRU layer and as additional information using concatenation.

\subsubsection{Evaluation metrics}
We evaluate the performance of our model and the baselines using a top-k recommendation setting. Accordingly, we use the following  common metrics:
$Precision@k$, $Hit@k$, and $NDCG@k$, with ranking cutoffs of $k\in\{1,2,5\}$. We note that, for $k=1$, all three metrics have the same value. Hence, for this case, we report only $Precision@1$. We report the average metrics over all recommendation tasks in the test-set. We validate statistical significance of the results using a two-tailed paired Student's t-test for 95\% confidence with a Bonferroni correction. 

%% file: 4-results.tex
\subsection{Main Results}
\begin{table}[]
\caption{\label{table:main}Main results. Boldfaced results indicate a statistically significant difference.}
\center\setlength\tabcolsep{3.0pt}
\small
\begin{tabular}{|l|ccc|cc|cc|} 
\hline
\multicolumn{1}{|c|}{\multirow{2}{*}{Model}} & \multicolumn{3}{c|}{Precision@k}                 & \multicolumn{2}{c|}{HIT@k}      & \multicolumn{2}{c|}{NDCG@k}      \\ 
\cline{2-8}
\multicolumn{1}{|c|}{}                       & @1             & @2             & @5             & @2             & @5             & @2             & @5              \\ 
\hline
\textbf{\textbf{\Ttd}}  (our model)     &        \textbf{43.51} &        \textbf{33.30} &        \textbf{21.56} &  \textbf{62.19} &  \textbf{80.85} &   \textbf{35.61} &   \textbf{26.12} \\
$\gru_C$    &        39.08 &        31.52 &        21.24 &  58.76 &  79.43 &   33.23 &   25.07 \\
\bst        &        38.15 &        31.21 &        21.28 &  58.09 &  79.57 &   32.78 &   24.94 \\
$\fdsa^-$   &        37.84 &        31.03 &        21.14 &  57.80 &  78.91 &   32.57 &   24.76 \\
$\fdsa^+$   &        36.95 &        30.37 &        21.01 &  56.57 &  78.35 &   31.86 &   24.46 \\
$\trans_T$  &        36.77 &        30.83 &        21.18 &  57.37 &  79.12 &   32.17 &   24.62 \\
\ca         &        36.06 &        30.50 &        21.09 &  56.78 &  78.70 &   31.76 &   24.41 \\
$\trans_A$  &        36.01 &        30.42 &        21.10 &  56.56 &  78.72 &   31.68 &   24.41 \\
$\gru_A$    &        35.71 &        30.41 &        21.08 &  56.58 &  78.65 &   31.61 &   24.35 \\
\lc         &        35.64 &        30.29 &        21.08 &  56.36 &  78.65 &   31.50 &   24.32 \\
\cdmf       &        32.71 &        28.15 &        20.41 &  52.10 &  75.30 &   29.18 &   23.15 \\
\seqfm      &        32.22 &        27.81 &        20.32 &  51.45 &  74.90 &   28.81 &   22.97 \\
\RSP        &        31.44 &        27.82 &        20.17 &  51.43 &  74.16 &   28.64 &   22.73 \\
\price (high to low)      &        15.51 &        15.69 &        15.57 &  27.96 &  52.60 &   15.65 &   15.60 \\
\price (low to high)      & 14.11          & 14.38          & 14.62          & 25.35          & 47.99          & 14.32          & 14.51 \\

\hline
\end{tabular}
\end{table}

We report the main results of our evaluation in Table~\ref{table:main}, where we compare our \textbf{\Ttd} model to all baselines. As can be observed, \textbf{\Ttd} outperforms all baselines over all metrics by a large margin. We next notice that, the best baselines after \textbf{\Ttd} are $\gru_C$ and \bst (Transformer-based). Common to both baselines is the fact that they apply a concatenation over the attribute embeddings. While there is no dimensional reduction in the attribute representation, there are still two drawbacks in such baselines. First, the semantic difference between the attributes is not kept; and therefore, the attention cannot be per attribute. Second, the number of parameters in these baselines drastically increases. \textbf{\Ttd}, on the other hand, 
requires only a slight expense of parameters.

Among the next best baselines, are those that employ a Transformer/Attention pooling (i.e., $\trans_T$, $\fdsa^-$, and $\fdsa^+$). Next to these are those that use average-pooling (i.e., $\trans_A$ and $\gru_A$). These empirical results demonstrate that, the attribute reduction mechanism is important to avoid losing important attribute information before entering the sequence block. Therefore, \textbf{\Ttd}, which does not require any such reduction in the attribute representations, results in overall best performance. 

Observing the difference between $\fdsa^+$ (as the only baseline that uses the original item-IDs) and $\fdsa^-$, further shows that using the original \emph{item-ID} resolves in worse results, as it causes an over-fit. This helps verifying the importance of representing items based on their (less sparse) attributes. Furthermore, it is worth noting that, using the \textbf{original} \emph{item-ID} as an attribute harms the performance of all other baselines as well.  Therefore, we implement all other baselines in this work (including our own model) without the original \emph{item-ID} attribute (yet, still with its hashed version).

\begin{figure*}
    \centering
    \subfloat[\# Blocks ($L$)]
        {\includegraphics[width=0.19\textwidth]{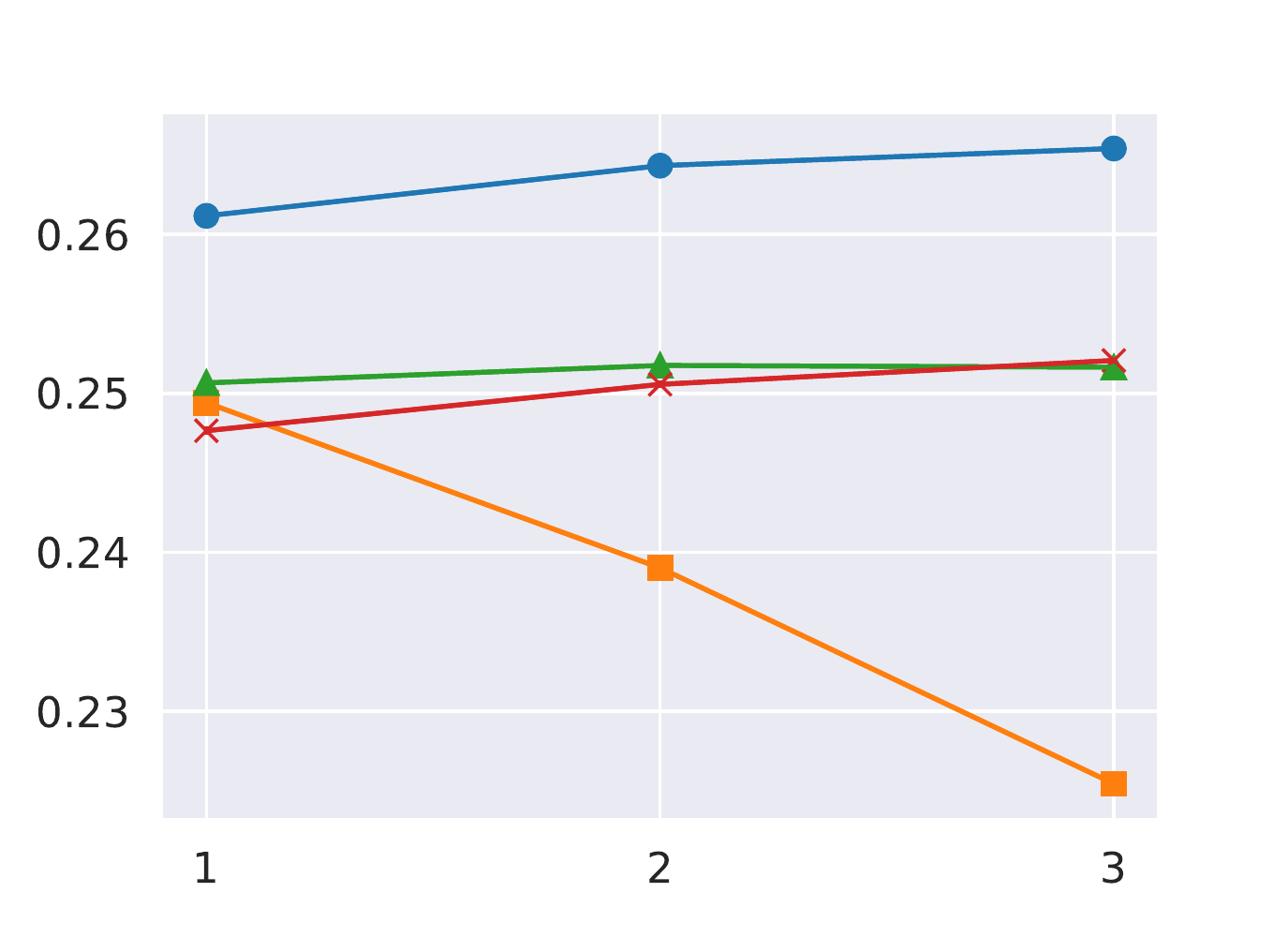}}
    \subfloat[\# Heads ($h$)]
    	{\includegraphics[width=0.19\textwidth]{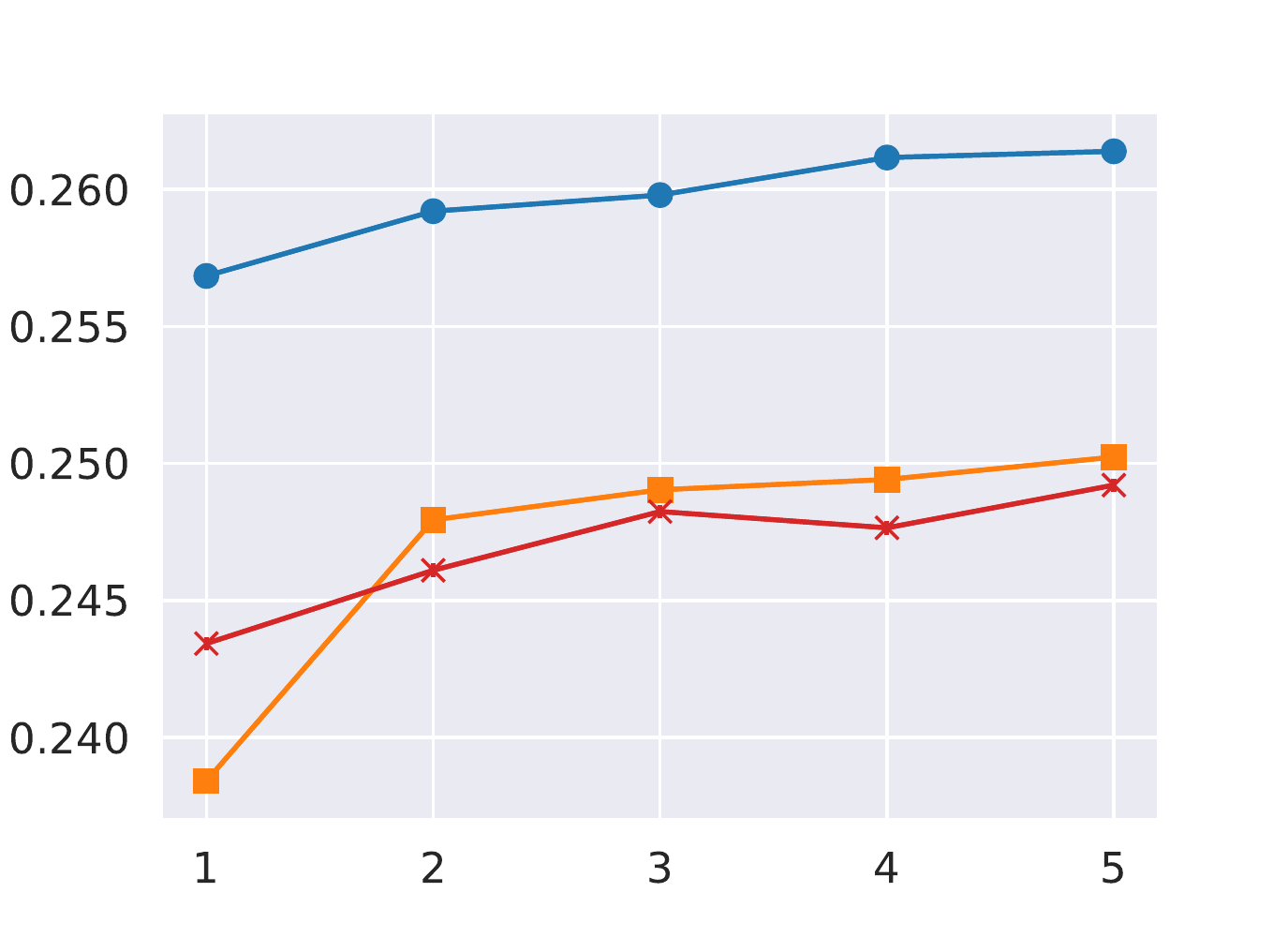}}
    \subfloat[Embedding Size ($d$)]
    	{\includegraphics[width=0.2\textwidth]{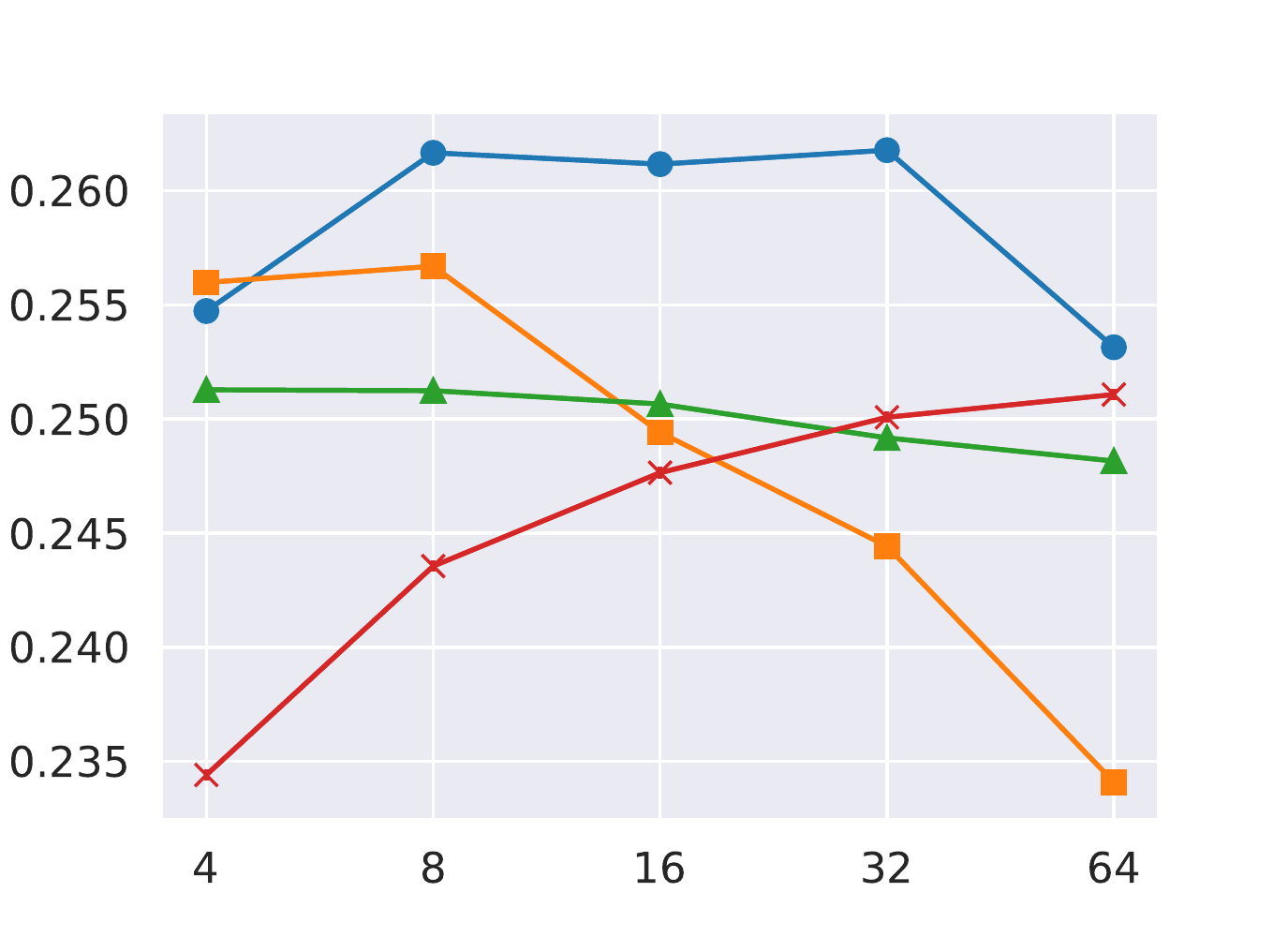}}
    \subfloat[Sequence Length ($N$)]
        {\includegraphics[width=0.2\textwidth]{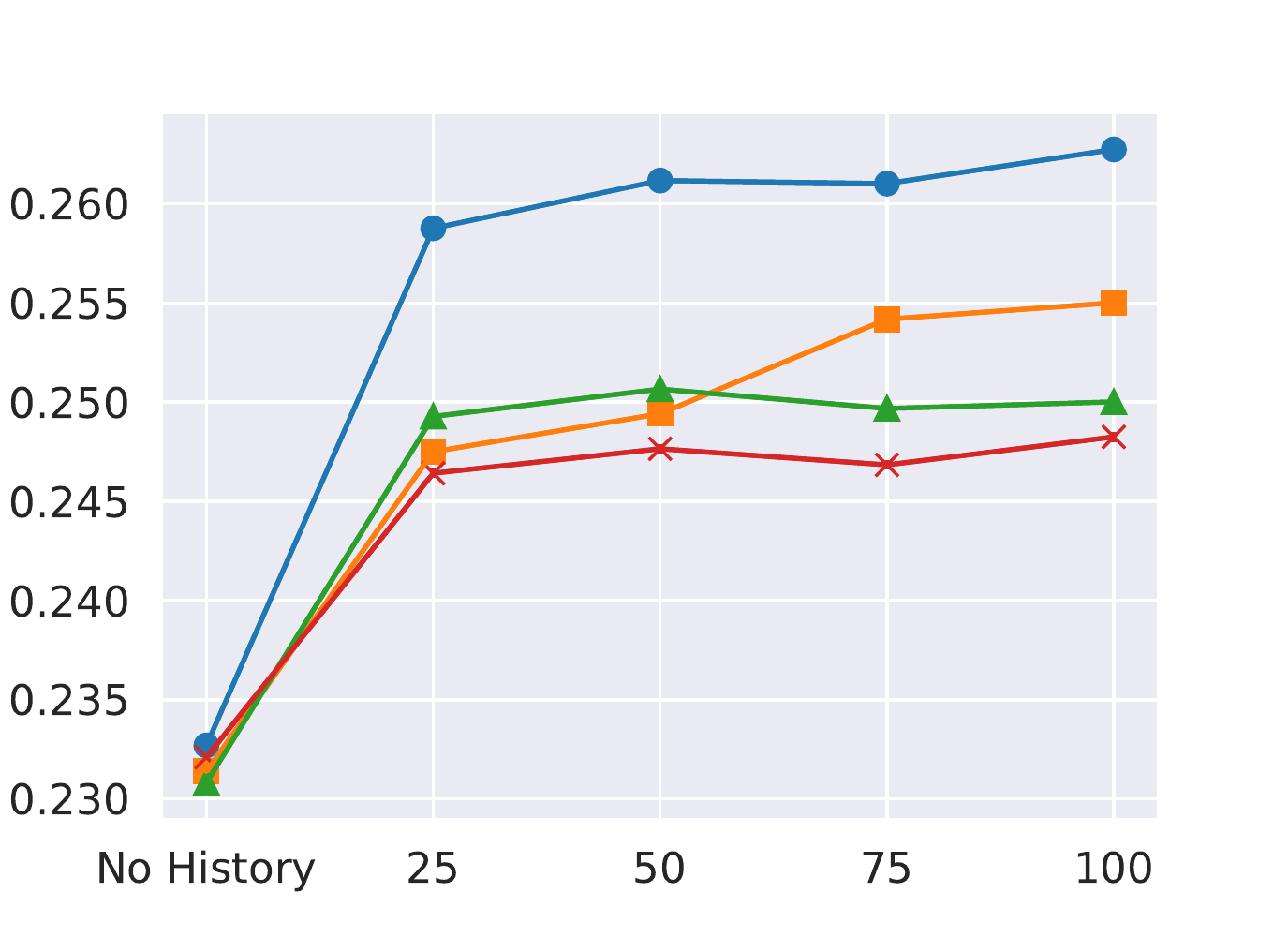}}
    \subfloat[Days Considered]
    	{\includegraphics[width=0.2\textwidth]{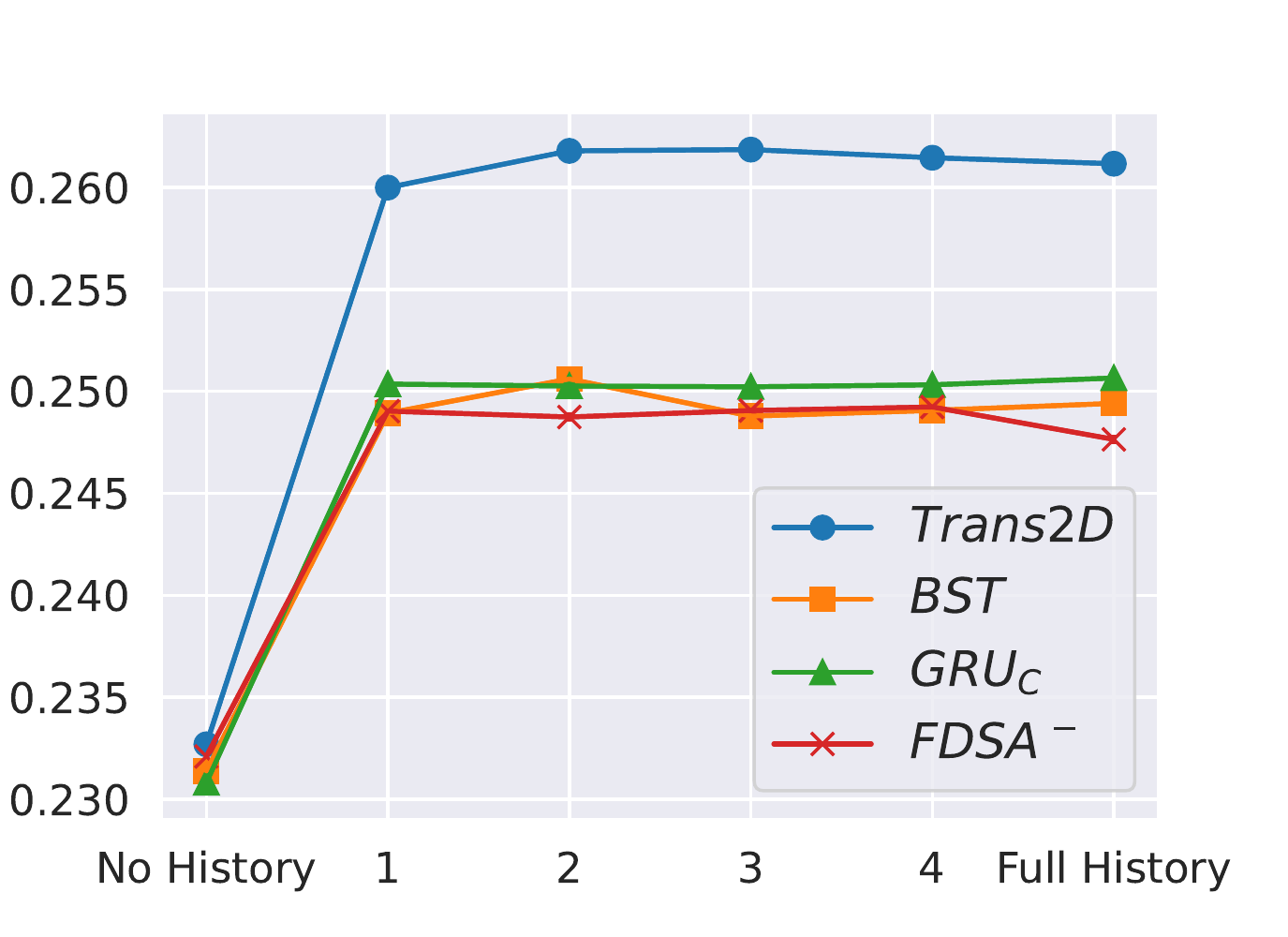}}
    \caption{\label{fig:sensitivity}Parameter Sensitivity results. Y-axis represents $NDCG@5$, while similar trends are observed for all other metrics.} 
\end{figure*}

We next observe that, the context-aware baselines (i.e., \ca and \lc) perform similar to the average-pooling baselines. Here we note that, these baselines also perform average-pooling over the context-features (also referred to as the ``next item features''). 
This supports our choice of adding the next possible item to the end of the sequence. Such a choice results in a similar effect to the context-aware approach that adds the item's features as context not before the sequence (GRU) block is applied. Moreover, since we have a relatively small recall-set in our setting, it is possible to concatenate the predicted item features to the history sequence and leverage the \textbf{\Atd} block to encode also the recall-set item attributes together with the sequence. 

Finally, we can observe that, ordering by the \RSP attribute results in a reasonable baseline on its own. This in comparison to both \price baselines that only provide a weak signal. This implies that, over our sampled users, the price does not pay a big deal, but rather the short-term interactions. This fact is further supported in our next parameter sensitivity analysis, see Figure~\ref{fig:sensitivity}(e). 

\subsection{Parameter Sensitivity}
\label{sec:parameter_sensitivity}


To better understand how the \textbf{\Ttd} model parameters influence its performance, we check its sensitivity to five essential parameters: 1) $L$, number of attention blocks, 2) $h$, number of heads used in the attention mechanism, 3) $d$, each attribute embedding size, 4) $N$, maximum sequence length, and 5) maximum number of days (back) to be considered in the sequence.
For each parameter sensitivity check, we set the rest to their default values (i.e., $L=1$, $h=4$, $d=16$, $N=50$, and days=`Full History' -- all days are being considered). 

We report the sensitivity analysis results (using NDCG@5) in Figure~\ref{fig:sensitivity}. For reference, we also report the sensitivity of the three next best baselines: $\gru_C$, $\bst$ and $\fdsa^{-}$. In general, we can observe that, in the majority of cases, our model outperforms the baselines. Furthermore, almost in all cases, the other baselines  behave similarly to our model. Per parameter, we further make the following specific observations about our model's sensitivity: 

\textbf{\# Blocks} ($L$): We notice that, adding a second \textbf{\Atd} block on top of the first helps to learn second-order dependencies. This makes sense, as the two attention matrices of the model, $\A^I$ and $\A^C$, can capture attention only in the same row or column. The only attention type that can capture dependencies between two cells not in the same row or column is $\A^F$. Yet, during the attention calculation on $\A^F$, the cells are not aware of the other values in their row or column. Therefore, a second block is able to capture such second-order dependencies 
(e.g., price comparison of two different appearances of the same item or leaf-category). 
We can further observe that, the third block has less influence on performance. 
Finally, 
similar to the results reported in~\cite{10.1145/3326937.3341261}, with additional blocks, $\bst$'s performance actually declines. 

\textbf{\# Heads} ($h$): Similar to normal attentions, utilizing more heads enables the attention mechanisms to learn multiple patterns. 
Here, we explore two main options for head sizes: 1) each head size equals the input size or 2) all head sizes together equal the input size. For our model and $\fdsa^{-}$ the first option is better, while for $\bst$, the second option is better. For the latter, this can be explained by its already over-sized input and data sparsity (see further details next in Embedding Size sensitivity). 
Overall, we observe that, $4$ heads resolves with a good performance, while $5$ heads already has a minor impact on our model's performance. 

\textbf{Embedding Size} ($d$): 
Interestingly, for our model, an embedding size of $d=32$ resolves with the best performance. While the trend for $d<32$ can be explained by too small representation sizes, the interesting trend is for $d>32$. This can be explained due the vocabulary size of each attribute. While most attributes include only few or tens of unique values, only \emph{user-ID}, \emph{level1-category}, and \emph{leaf-category} hold $40,344$, $1,293$, and $25,449$ unique values, respectively. Therefore, while a larger representation may help the latter, the rest of the attributes suffer from over-representation and sparsity. We note that, those baselines that use concatenation ($\gru_C$ and \bst) can be thought as using a larger representation for each item. This fact supports the performance decline we observe for these baselines: with larger representations, more over-fit. Differently from the former two, $\fdsa^{-}$ actually manages to improve (up to some point)  with the increase in embedding size. We attribute this to its usage of the vanilla attention pooling over the attribute embeddings, which in turn, allows it to represent attributes together in an efficient way, even if some are very sparse.


\textbf{Sequence Length} ($N$) + \textbf{Days Considered}:  The result of the \RSP baseline implies that, users' \wl priorities are usually driven by short-term patterns. Such patterns are captured by users' recently viewed item pages. We can observe that, 
for both parameters, $N=50$ or $days=1$ is enough in order to capture most of the important short-term signal. We further see a small improvement in extending the time range or sequence length. This implies that there is still a weak signal that can be captured. 
Since the sequence model is a Transformer, there is no memory-gate involved (like in GRU); hence, these short-term patterns represent a real user-behaviour.

\subsection{Ablation Study}

We next perform an ablation study and report the results in Table~\ref{table:ablation}. To this end, every time, we remove a single component from the \textbf{\Ttd} model and measure the impact on its performance. We explore a diverse set of ablations, as follows: 

$-\Ltd$: We switch every \Ltd layer with a \Lod layer, meaning all channels receive the exact same weights. This ablation allows to  better understand the importance of the attribute-alignment.
While this modified component downgrades performance, it still outperforms all baselines while keeping the exact same number of parameters as in a regular attention layer (except for the three $\alpha$ scalars used in Eq.~\ref{eq:final_attn}). This confirms that, the improvement of the full \textbf{\Ttd} results are not only driven by the \Ltd layer but also by the attention mechanism itself.

$-\A^F$,$-\A^C$,$-\A^I$: We remove each one of the three attention parts from the final attention calculation in Eq.~\ref{eq:final_attn}. These ablations allow to better understand the importance of each attention type to the full model implementation.
The most important attention part is $\A^I$, 
showing that the relationship between user's items is the most important, following the motivation of the 1D-attention mechanism.
$\A^I$ is followed by $\A^C$ and then by $\A^F$. As we can further observe,  $\A^F$ does not hold significant importance. We hypothesis that, while $\A^F$ is the only attention part that can capture dependencies between two cells not in the same row or column, during the attention calculation on $\A^F$, the cells are not aware of the other values in their row or column. This can be over-passed by adding an additional \textbf{\Atd} layer, as shown and discussed in Section~\ref{sec:parameter_sensitivity} over the \# Blocks sensitivity check, see again Figure~\ref{fig:sensitivity}(a).

$-RVI,-\wl$: We remove from the history sequence all items having \emph{interaction type}=RVI (item-page-view) and \emph{interaction type}=\wl (click), respectively. These ablations allow to understand the importance of each \emph{interaction type} to the WLR task.
While both are important, \wl clicks hold a stronger signal. This actually makes sense: while item page views (RVIs) provide only implicit feedback about user's potential interests, 
clicks provide explicit feedback that reflect actual user \wl priorities. 

\begin{table}[]
\caption{\label{table:ablation}Ablation results. Starting from the second row, on each, a single component is removed from the model. }
\center\setlength\tabcolsep{3.0pt}
\small
\begin{tabular}{|l|ccc|cc|cc|} 
\hline
\multicolumn{1}{|c|}{\multirow{2}{*}{Model ~}}        & \multicolumn{3}{c|}{Precision@k}                 & \multicolumn{2}{c|}{HIT@k}      & \multicolumn{2}{c|}{NDCG@k}      \\ 
\cline{2-8}
\multicolumn{1}{|c|}{}                                & @1             & @2             & @5             & @2             & @5             & @2             & @5              \\ 
\hline
\textbf{\textbf{\Ttd}}  (our model)     &        43.51 &        33.30 &        21.56 &  62.19 &  80.85 &   35.61 &   26.12 \\
-$A^F$            &        44.01 &        33.42 &        21.56 &  62.42 &  80.85 &   35.82 &   26.20 \\
-time             &        43.25 &        33.17 &        21.55 &  61.93 &  80.79 &   35.45 &   26.06 \\
-RVI              &        43.21 &        33.09 &        21.51 &  61.81 &  80.62 &   35.38 &   26.01 \\
-$A^C$            &        43.03 &        33.06 &        21.53 &  61.72 &  80.69 &   35.32 &   26.00 \\
-item             &        42.21 &        32.64 &        21.40 &  60.89 &  80.13 &   34.80 &   25.74 \\
-Linear2D         &        41.93 &        32.71 &        21.50 &  61.02 &  80.56 &   34.80 &   25.79 \\
-\wl              &        40.07 &        31.48 &        21.08 &  58.56 &  78.49 &   33.42 &   25.08 \\
-$A^I$            &        38.83 &        31.60 &        21.36 &  58.84 &  79.92 &   33.24 &   25.14 \\
-history          &        33.04 &        28.45 &        20.46 &  52.66 &  75.53 &   29.49 &   23.27 \\
-position         &        32.18 &        26.77 &        19.86 &  49.45 &  72.81 &   27.99 &   22.51 \\
\hline
\end{tabular}
\end{table}


$-time,-item,-position$: For each, we remove a group of attributes in order to better understand the importance of the attributes to the WLR task. For $-time$, we remove the \emph{hour}, \emph{day}, and \emph{weekday} attributes. For $-item$, we remove the \emph{price}, \emph{condition}, \emph{level1-category}, \emph{leaf-category}, and \emph{sale-type} attributes. For $-position$, we remove the \emph{position-ID}, \emph{relative-snapshot-position} (\RSP), \emph{snapshot-ID}, \emph{hash-item-ID}, and \emph{hash-seller-ID} attributes.
Examining the ablations results, shows that the most important attribute-set is the position-set, followed by the item-set, and finally by the time-set. Noticing that the position-set includes \RSP explains their high importance. The time-set is not that important as relative-time is encoded in the position-set. Additionally, the dataset spans over a relatively short period, resolving with less data to learn the importance of a specific weekday, hour of the day, or day of the month.

$-history$: To understand the importance of the entire history, we remove all the history sequence, and solely predict over $\WL_t$ -- the last snapshot.
We see that, not observing any history, drastically harms the performance of the model. This strengthens the argument that considering user's history is important for the WLR task. 

\begin{figure*}
\centering
\includegraphics[width=0.8\textwidth]{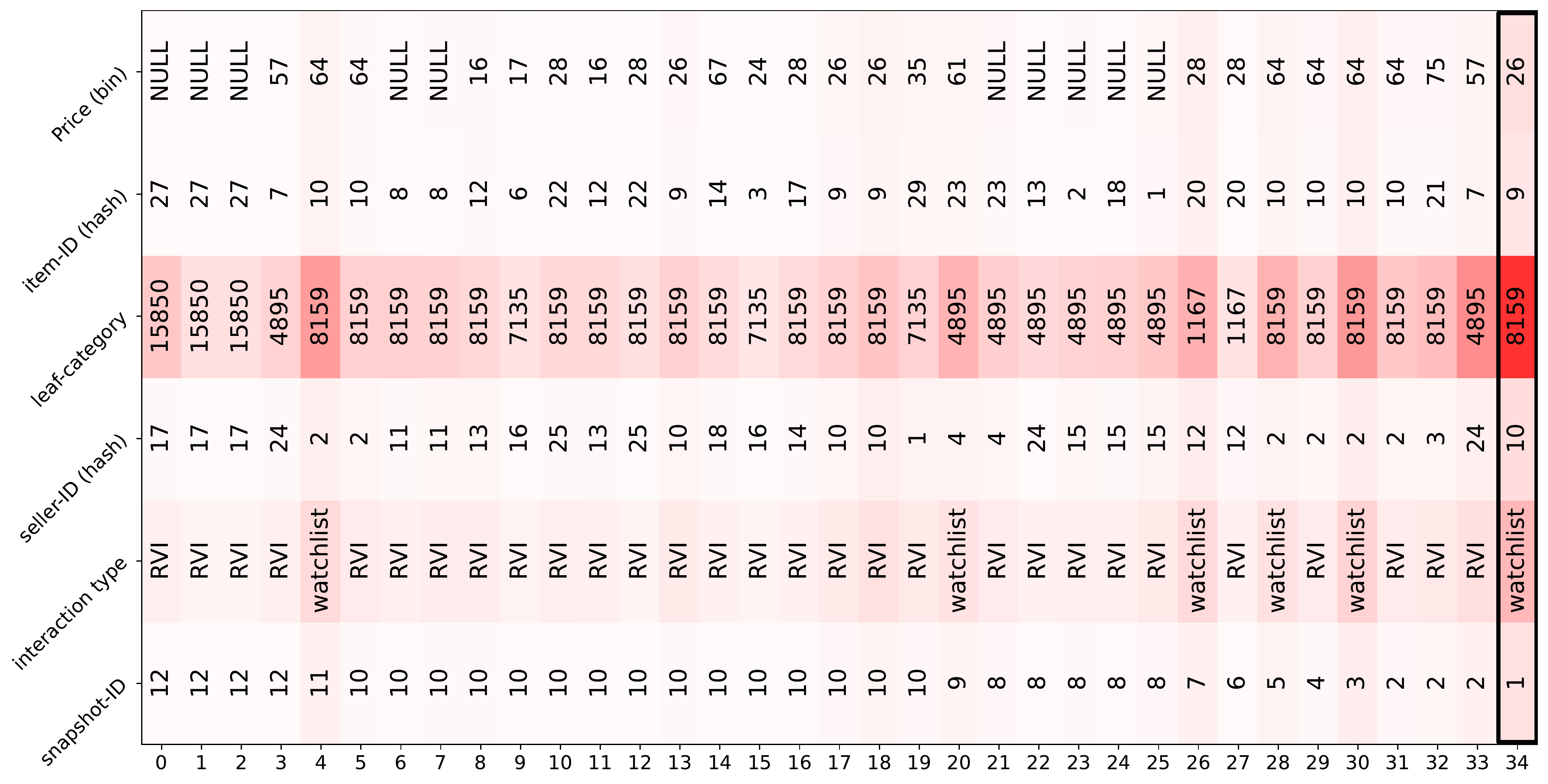}
\caption{Qualitative Example. For brevity, only 6 out of the 16 possible attributes are presented. The darker the item-attribute cell color is, the more important it is. 
The current candidate is surrounded with the black rectangular. 
}
\label{fig:qualitative}
\end{figure*}

\subsection{Qualitative Example}

We end this section with a qualitative example (see Figure~\ref{fig:qualitative}) that visualizes the attention values assigned by the \textbf{\Ttd} model to items' attributes in a given user-sequence. The last item in the sequence is the \wl item for which the model makes a prediction. This item is actually clicked by the user ($y=1$) and the model's prediction is aligned, assigning it the highest likelihood amongst all \wl snapshot items ($\hat{y}=0.697$).

As the learned attention is a 4D-array, representing the attention between any two (item,attribute) pairs (see Section~\ref{sec:tensordot}), we first pick the attention over the last item in the sequence, and then average over the channel dimension. Doing so, resolves us with a 2D-array that can be visualized to better understand the importance of each (item,attribute) pair on the prediction of the item.  Observing the channel dimension, we can notice that, different attributes have different importance. This strengths the importance of a 2D-attention that is able to learn the attention over a second axis.

As we can observe, given the next item to predict, the model emphasizes either previously clicked items related to the current predicted item (e.g., leaf-categories: 8159--`\texttt{Home Security:Safes}`, 4895--`\texttt{Safety:Life Jackets \& Preservers}`) or RVIs that belong to the same leaf-category of the current item (8159). Interestingly, the model emphasizes the prices of clicked items in the same leaf-category (8159), which are higher than the predicted item's price.

%% file: 5-conclusions.tex
\section{Summary and Future Work}
\label{sec:conclusions}
In this work, we presented a novel \wl recommendation (WLR) task. The WLR task is a specialized sequential-recommendation task, which requires to consider a multitude of dynamic item attributes during both training and prediction. To handle this complex task, we proposed \Ttd\ -- an extended Transformer model with a novel self-attention mechanism that is capable of attending on 2D-array data (item-attribute) inputs. Our empirical evaluation has clearly demonstrated the superiority of \Ttd. \Ttd  allows to learn (and preserve) complex user preference patterns in a given sequence up to the prediction time. 

Our work can be extended in two main directions. First, we wish to explore additional feedback sources such as historical user search-queries or purchases. Second, recognizing that \Ttd can be generally reused, we wish to evaluate it over other sequential recommendation tasks and domains.